\documentclass{aa}
\usepackage{graphics,epsfig}

\newcommand{\kms}{km~s$^{-1}$}

\newcommand{\acc}{~cm$^{-2}$}

\begin{document}
%\thesaurus{11(11.05.2; 11.06.1; 11.09.4), 12(12.03.3), 13(13.19.3)}

\title{A dwarf galaxy with a giant HI disk}
\titlerunning{A dwarf galaxy with a giant HI disk}
\author{Ayesha Begum\inst{1}\thanks{ayesha@ncra.tifr.res.in},
        %Ayesha Begum\inst{1}
        Jayaram N. Chengalur\inst{1} and
%        Jayaram N. Chengalur\inst{1}\thanks{chengalur@ncra.tifr.res.in}
%	I. D. Karachentsev\inst{2}\thanks{ikar@luna.sao.ru}
%	M. E. Sharina\inst{2}\thanks{sme@sao.ru}
	I. D. Karachentsev\inst{2} 
%	M. E. Sharina\inst{2}
}
\authorrunning{Begum et al.}
\institute{1-National Centre for Radio Astrophysics,
Post Bag 3, Ganeshkhind, Pune 411 007,
\\
2-Special Astrophysical Observatory, Nizhnii Arkhys 369167, Russia}
\date{Received mmddyy/ accepted mmddyy}
\offprints{Ayesha Begum}
\abstract{
We present Giant Meterwave Radio Telescope (GMRT) HI 21cm images of a nearby dwarf 
irregular galaxy NGC 3741 ($M_B \sim -13.13$) which show it to have a gas 
disk that extends to $\sim 8.3$ times its Holmberg radius. This makes 
it probably the most extended gas disk known. Our observations 
allow us to derive the rotation curve (which is flat in the outer regions)
out to $\sim$38 optical scale lengths. NGC~3741 has a dynamical mass to light 
ratio of $\sim$107 and is one of the ``darkest'' irregular galaxies known.
However, the bulk of the baryonic mass in NGC~3741 is in the form of gas 
and the ratio of the dynamic mass to the baryonic mass ($\sim 8$), falls 
within the range that is typical for galaxies. Thus the dark matter halo
of NGC~3741 has acquired its fair share of baryons, but for some reason, 
these baryons have been unable to collapse to form stars. A comparison
of NGC~3741's dark halo properties with those of a sample of galaxies with
well measured rotation curves suggests that if one has to reconcile the
observations with the expectation that low mass galaxies suffer fractionally
greater baryon loss then baryon loss from halos occurs in such a way
that, in the net, the remaining baryons occupy a fractionally smaller
volume of the total halo.
}
\maketitle

\section{Introduction}
\label{intro}

    Numerical simulations of hierarchical galaxy formation predict 
that galaxies should  have dark halos with virial radius $\sim 8-10$ 
times the size of the stellar disk and a total mass 
of $\sim 50-70$ times the stellar mass. Since the HI disks of 
galaxies typically extend to only $\sim 2$  times the optical radius, 
one cannot test these model predictions using HI rotation curves. 
The mass distribution on such large scales can however be probed 
using weak lensing or the kinematics of faint satellite galaxies. 
Though neither of these techniques can be applied to  individual 
galaxies, when applied to a large sample of galaxies, they do 
provide at least qualitative confirmation (for $\rm{L_*}$ galaxies) 
of the models (see e.g. Brainerd, 2004).

   While, dark matter halos are expected to be self similar (Navarro et al. 1997), 
there are no direct determinations of the typical virial size and mass of dwarf 
galaxy halos. For small galaxies, both weak lensing as well as
the kinematics of still fainter companions are correspondingly
difficult to measure. Further, the rotation curves for most of the 
faint dwarf irregulars are typically rising even at the last measured 
point, implying that one has been unable to probe beyond the core
of the dark halo, leave alone its virial radius. There are however 
a few dwarf galaxies known with unusually extended HI disks (e.g. DDO154, 
Carignan \& Beaulieu, 1989; NGC 2915, Meurer et al. 1996;ES0215-G?009, 
Warren et al. 2004), where the HI extends to more than 5 times the 
Holmberg radius (R$_{\rm{Ho}}$).  The extended HI gas in 
such galaxies traces the dark matter potential upto large galacto-centric
radii and hence provide a unique opportunity to measure the larger scale
mass distribution around dwarf galaxies. We discuss here  GMRT 
\footnote{The GMRT is operated by the National Center for Radio Astrophysics 
of the Tata Institute of Fundamental Research} 
HI observations of one such galaxy with an extended HI disk,  viz.
NGC~3741. Our GMRT observations show that this galaxy has a regular HI
disk that extends out to  $\sim 8.3 $ times R$_{\rm{Ho}}$. Like other
known galaxies with unusually extended disks, NGC 3741 is also located in the
vicinity of the local group. 
 
\section{Observations and data reduction }
\label{sec:obs}     

\begin{figure*}[t!]
\epsfig{file=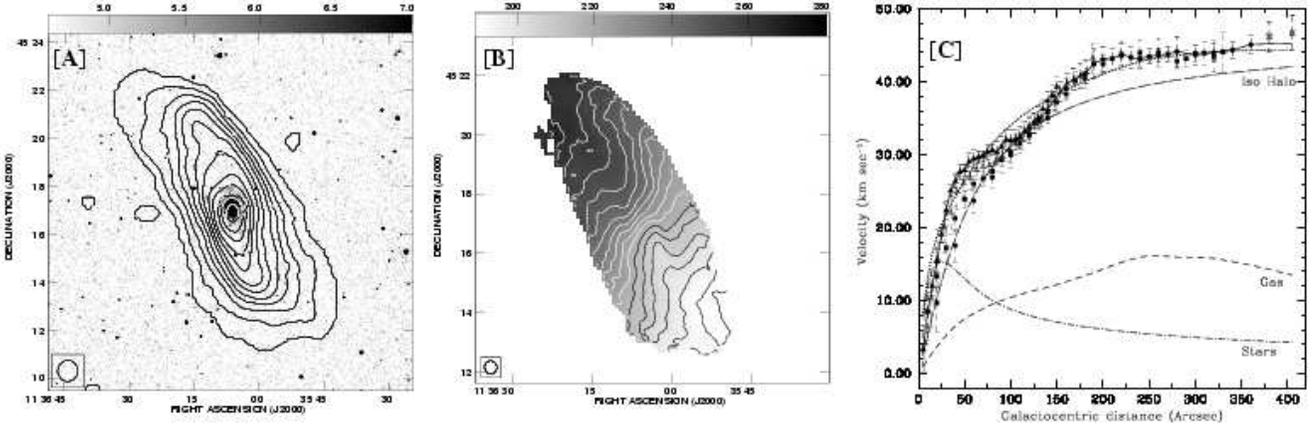,width=7.0truein,height=2.3in}
\caption{{\bf{[A]}}The B band optical image of NGC 3741 (greyscales) with
         the GMRT 52$^{''}\times49^{''}$  resolution integrated HI
         emission map (contours) overlayed. The contour levels are 0.1, 1.3,
        2.6, 3.8, 5.0, 6.5, 7.8, 9.6, 11.3, 14.4, 17.0, 20.5 $\times 10^{20}$
        atoms cm$^{-2}$ {\bf{[B]}} The HI velocity
        field of NGC 3741 at 26$^{''}\times 24^{''}$ resolution. The
        contours are in the steps of 5 \kms and range from 185.0~\kms
        to 265.0~\kms.{\bf{[C]}}  The rotation curve for  NGC 3741
        derived from the intensity weighted velocity field at 52$^{''}
        \times 49^{''}$, 42$^{''}\times 38^{''}$,  26$^{''}\times 
        24^{''}$, 19$^{''}\times 15^{''}$ and 13$^{''}\times 11^{''}$
        resolution shown as stars, circles, squares, open triangles and
        filled triangles respectively. The best fit mass model using an
        isothermal halo is  shown as dotted line.
}
\label{fig:mom}
\end{figure*}

    The GMRT observations of dwarf irregular galaxy NGC~3741 ($\rm{M_B} \sim -13.13$)
 	\footnote{Using apparent blue magnitude 14.38$\pm$0.1 (Bremnes et al. 2000),
                  galactic extinction correction Ab = 0.11 mag (Schlegel et al.)
                  and TRGB distance of 3.0~Mpc (Karachentsev et al. 2004).}
were conducted on 22 July and 26 Aug 2004.
The channel spacing was $\sim 1.7$~\kms, and the total velocity coverage was $\sim 230$ \kms.
3C286 was used as the flux and bandpass calibrator, while the VLA calibrator 1227+365
was used to do phase calibration. The total on-source time was $\sim$ 8 hours.

     The data were reduced using standard tasks in classic AIPS.  Image data cubes were 
made at various resolutions (see the caption of Fig.~\ref{fig:mom}).
%including 52$^{''}\times 49^{''}$, 42$^{''}\times 38^{''}$, 
%26$^{''}\times 24^{''}$, 19$^{''}\times 15^{''}$ and 13$^{''}\times 11^{''}$.
%The RMS noise per channel at these resolutions is 3.8~mJy, 3.2 mJy, 2.9 mJy, 
%1.5 mJy and  2.0 mJy respectively. 
Moment maps  of the line emission were made 
using the AIPS task MOMNT. 

\section{Results and Discussion}
\label{sec:res}
\subsection{HI distribution and kinematics}
\label{ssec:HI_dis}

 Fig ~\ref{fig:mom}[A] shows the integrated HI emission from NGC~3741 (at 
52$''\times49''$ resolution), overlayed on the digitised sky survey image. 
As can be seen, the HI distribution of NGC 3741 is regular and extends 
to $\sim$ 13.8$^{\prime}$ in diameter, at a level of $\sim 1 \times 10^ {19}$ 
cm$^{-2}$ ($\sim$ 13.4$^{\prime}$ in diameter, at a level of $\sim 5 \times 10^ {19}$
cm$^{-2}$). The  Holmberg diameter of NGC 3741 measured from the B band 
surface brightness profile is $\sim100^{\prime\prime}$ (Bremnes et al. 2000) 
$-$ the HI disk is hence $\sim$ 8.3 times the Holmberg diameter. 
The HI distribution has a central bar, more clearly seen in the higher
resolution maps (not shown), which is approximately coincident with
the optical emission.

The integrated flux estimated from the HI emission profile is $74.7\pm7.5$~Jy~\kms
(corresponding to an HI mass of $1.6\pm0.4 \times{10}^{8} \rm{M_\odot}$).
The HI flux measured at GMRT is more than the single dish flux (
53.0~Jy~\kms, ~\cite{schneider}). The single dish observations,
done with the Greenbank 300ft telescope with a single pointing centered
on the optical galaxy, would have underestimated the total flux of
the galaxy, as the extent of the HI disk of the galaxy is
bigger than the beam size of the telescope. The flux integral estimated from
our GMRT observations is also likely  to be a lower limit on the total HI flux
of the galaxy as the GMRT is well known to resolve out flux for galaxies that
are this extended (e.g. Omar, A., 2004). Our flux measurement hence places
a lower limit on the $\rm{M_{\rm{HI}}/L_{\rm{B}}}$ ratio of =$5.8\pm1.4$.

         The velocity field of NGC 3741 (at 26$''\times 24''$ resolution)
is shown in Fig.~\ref{fig:mom}[B]. The velocity field is regular, with a warp
being seen in the outer regions of the galaxy, as well as distortions in the
central region, which correlate with the central bar. The central distortions
of the isovelocity contours are more prominent in the higher resolution
velocity fields.

     Fig.\ref{fig:mom}[C] shows the rotation curves for NGC 3741 derived 
from the various resolution velocity fields using  the tilted ring model.
The kinematical inclination of the galaxy was found to vary from $\sim$ 58$^\circ$ 
to $70^\circ$ across the galaxy, whereas the kinematical position angle 
varied from $\sim$ 33$^\circ$ to 47$^\circ$. The final adopted rotation curve 
(solid line) is measured upto $\sim$ 38 disk scale length (the B band scale length 
is 10.75$^{\prime\prime}$; Bremnes et al. 2000). The rotation curve shows a 
flattening beyond $\sim 200^{\prime\prime}$;  NGC~3741 is  one of  
the faintest known dwarf galaxies to show a clear flattening of 
the rotation curve. A steep rise in the rotation velocities 
within the Holmberg radius is probably related to the bar at 
the center of the galaxy. The rotation curve was also derived 
(at each spatial resolution) independently for the approaching and 
the receding side of the galaxy. These curves match within the errorbars. 
Correction for the ``asymmetric drift'' was also done and was found to be small 
compared to the errorbars at all radii.

    Modified isothermal and NFW mass models were fit to derived rotation curve.
Fig.~\ref{fig:mom}[C] shows the best fit modified isothermal
halo model, which has a mass to light ratio  ($\Gamma_B$) for the stellar disk 
(which was assumed to be an exponential disk with intrinsic thickness ratio of 0.25)
of 0.9$\pm$0.1, a halo with core radius r$_{\rm{c}}$=0.7$\pm$0.1 kpc and core density 
$\rho_0=77.9\pm8.9~\times10^{-3} \rm{M_\odot}$ pc$^{-3}$. The best fit NFW mass model (not
shown) gives $\Gamma_B$ of 0.3$\pm$0.1, a concentration parameter c=11.4$\pm$0.8 and 
V$_{200}=35.3\pm1.4$ \kms. Both type of dark halos provide a comparable fit to the 
derived rotation curve. The bar linked central distortion of the velocity field
is a major contributor to our inability to distinguish between these two models.

\subsection{Discussion}
\label{ssec:discuss}

\begin{figure*}[t!]
\epsfig{file=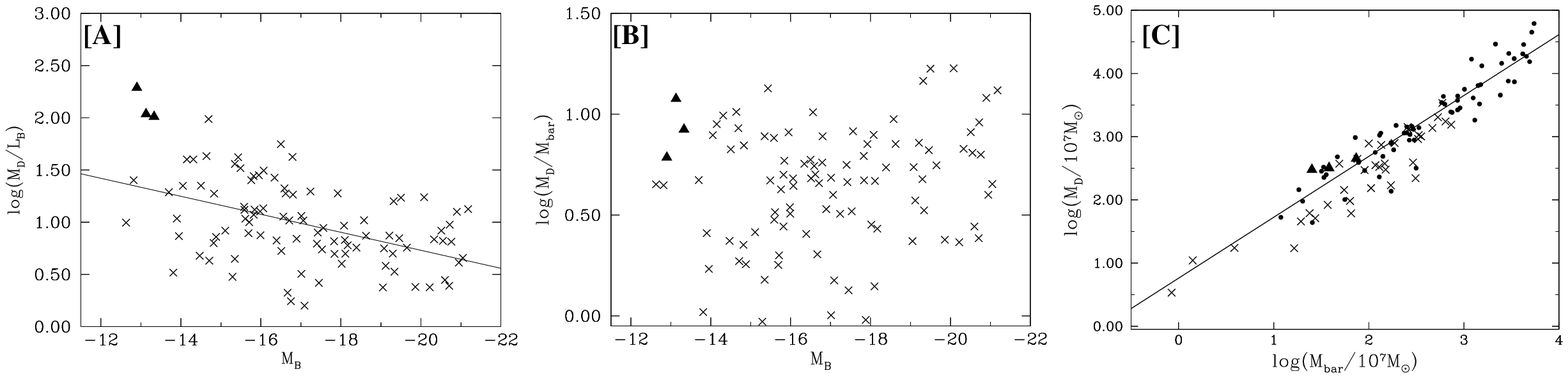,width=7.0truein}
\caption{
  {\bf [A]} $\rm{M_{D}/L_{B}}$ as a function
            of M$_{\rm{B}}$ for a sample of galaxies with HI rotation curves from
            Verheijen(2001), Swaters (1999) and Cote et al. (2000). Galaxies
            with very extended HI disks viz. NGC~3741, DDO~154 and~ESO 215-G?009 are marked
            as triangles. The solid line shows the best fit relation.
  {\bf [B]} $\rm{M_{D}/M_{\rm bar}}$ as a function
            of M$_{\rm{B}}$  for the same sample of galaxies.
  {\bf [C]} M$_{\rm{D}}$ as a function of $M_{\rm{bar}}$ for the same sample of galaxies.
            Galaxies with rising rotation curves are
            marked as crosses while galaxies with flat rotation curves
            shown as filled circles. The solid line is the best fit relation for galaxies with
            flattened rotation curves.
}
\label{fig:corr}
\end{figure*}

         NGC~3741 probably has the most extended HI disk known. In comparison, 
VLA observations of DDO154, (the commonly cited example of the largest known HI disk), 
detected HI up to $\sim$ 5 times R$_{\rm{Ho}}$  (Carignan \& Beaulieu, 1989); the follow-up
Arecibo observations detected HI upto $\sim$ 8.5 times R$_{\rm{Ho}}$  (Hoffman et al. 1993),
although these last observations do not lead to a good quality rotation curve. Most 
other galaxies with extended HI envelopes show severe tidal distortions, e.g. 
IC~10 (Wilcot \& Miller 1998).

         The total dynamical mass of NGC~3741 (at the last measured point of the rotation 
curve) is $3.0\pm0.4~\times10^9{\rm{M_\odot}}$, i.e. the dynamical mass to light ratio is
$\rm{M_{D}/L_B}=107\pm16.4$; NGC~3741 is hence one of the ``darkest'' irregular galaxies 
known. A fundamental question which arises is whether NGC~3741's dark halo is the same as
those of similar dwarf galaxies with less extended disks, i.e. is the extended HI just 
a fortuitous tracer of the gravitational potential, or does the presence of such an 
extended disk mean that we are dealing with a different (e.g. more massive than typical)
halo? Equivalently, should we regard NGC~3741 as a 13th magnitude dwarf which somehow 
acquired a lot of gas, or should we regard it as a galaxy that should ``rightly'' have
been much brighter, but for some reason it failed to convert its gas
into stars? Indeed, why do some galaxies have such extended HI disks?

  We start by trying to address the first set of questions, viz. regarding the nature
of the dark halo of NGC~3741 as compared to those of ``similar'' galaxies. In 
Fig.~\ref{fig:corr}[A] is shown the dynamical mass to light ratio M$_{\rm D}$/L$_{\rm B}$ 
(at the last measured point of the rotation curve) as function of the absolute 
blue magnitude (M$_{\rm B}$) for a sample of galaxies with measured HI rotation curves.
Galaxies with the most extended HI disks viz. NGC~3741, DDO~154 and 
ESO~215-G?009
\footnote{ESO~215-G?009  has a 
                      very uncertain $\rm{M_{D}/L_{B}}$. In 
                      this paper we use the nominal value suggested by Warren 
                      et al. (2004).} 
are marked with triangles. As can be seen, the well known (albeit weak and noisy)
trend of increasing dynamical mass  to light ratio with decreasing galaxy luminosity 
is recovered, and in addition, dwarfs with an extended HI disks can be seen to 
have fairly extreme mass to light ratios. However, given that for galaxies with 
extended HI disks the total baryonic mass is dominated by the HI mass, instead of 
comparing $\rm{M_{D}/L_{B}}$ it might be more fair to compare dynamical to baryonic mass ratio
($\rm{M_{D}/M_{bar}}$), where 
M$\rm{_{bar}}$ is the baryonic mass. We define M$_{\rm{bar}}$=M$_{\rm{gas}}$+M$_*$,
and  assume a stellar mass to light ratio, $\rm{\Gamma_B}$=1.0 for all galaxies in the
sample. M$_{\rm{gas}}$ is defined as $1.4 \times $M$_{\rm HI}$. From 
Fig.~\ref{fig:corr}[B] one can see that $\rm{M_{D}/M_{bar}}$ does not show any 
correlation with M$_{\rm{B}}$, i.e. (proviso the naiveness of our baryonic mass 
calculation), even though dwarf galaxies tend to have a higher total mass to 
luminosity ratio, they have dynamical to baryonic mass ratios that are similar to
those of $\rm{L_*}$ galaxies. Further the dynamical mass to baryonic mass ratio for
galaxies with extended HI disks, though large, lies within the range of what
is observed for other galaxies in the sample. In particular, given its 
dynamical mass, NGC3741 does not have an anomalously small fraction of baryons.
The other interesting feature in Fig.~\ref{fig:corr}[B] is the uniformly 
large scatter in the dynamical to baryonic mass ratio and the lack of a 
statistically significant trend in M$_{\rm D}$/M$_{\rm bar}$ with galaxy size. 
Quantitatively, the average M$_{\rm bar}$/M$_{\rm D}$ is $0.30 \pm 0.03$ 
for galaxies brighter than $-17$~mag, and $0.28 \pm 0.03$ for galaxies fainter 
than $-17$~mag. This is in apparent contradiction to simulations of galaxy 
formation, which predict that dwarf galaxies should have both a smaller 
baryon fraction than large galaxies, and also a larger dispersion in the 
baryon fraction (Gnedin et al. 2002). These differences
in the baryon fractions of large and small galaxies come about because 
for small halos the heating of gas at reionization inhibits the capture 
of baryons and further a good fraction of the captured baryons are subsequently
lost due to the energy input from star formation.
However it is important to distinguish between the baryonic fraction as 
calculated in numerical simulations (which is generally defined as the 
ratio of the total mass within the virial radius to the total baryonic mass) 
and that plotted in Fig.~\ref{fig:corr}[B] which is the ratio of total 
dynamical mass to the total baryonic mass {\it within the radius to which 
the HI disk extends}.  Since baryons cool and collect at the centers of dark
matter halos, in any given halo the baryon fraction is a decreasing function 
of radius. This means that  one could reconcile 
the rotation curve data with the theoretical models
by requiring that the baryons in dwarf galaxies occupy a proportionately 
smaller fraction of their dark matter halos as compared to large galaxies. 
If this is what indeed happens, it is somewhat surprising that things 
nonetheless conspire  to give  a more or less constant baryon fraction within
the last measured point of the rotation curve. A similarly surprising 
correlation between baryonic mass and velocity width (the ``baryonic 
Tully-Fisher'' relationship) has been reported by Mcgaugh et al (2000).  
Of course, since many dwarf galaxies have rising rotation curves (i.e.
the gas and stars lie entirely within the cores of their dark matter halos)
it seems intuitive to suppose that the baryons in these galaxies occupy 
a smaller fraction of their dark matter halos. Since all the galaxies in 
our sample have measured rotation curves, it is straight forward to compare
the baryon fraction in galaxies with rising rotation curves with that in 
galaxies which have flat rotation curves. In Fig.~\ref{fig:corr}[C] 
M$_{\rm D}$ is plotted as a function of M$_{\rm bar}$, separately for 
galaxies with flat rotation curves (filled circles) and rising rotation curves 
(crosses). As can be seen, there is a substantial range of baryonic
masses at which one finds both galaxies with rising rotation curves  as 
well as galaxies with flat rotation curves. However, at any 
given baryon mass, galaxies with rising rotation curves tend to have a 
lower dynamical mass than galaxies with flat curves. Equivalently,
within the volume occupied by the baryons, galaxies with flat rotation
curves have smaller baryon to dynamical mass fractions than galaxies with
rising rotation curves. Quantitatively, the mean M$_{\rm bar}$/M$_{\rm D}$ 
is $0.25 \pm 0.02$ for galaxies with flat rotation curves, and 
$0.36 \pm 0.03$ for galaxies with rising rotation curves.

In essence, the baryon fraction within the last measured point 
of the rotation curve depends on both the baryon loss as well as
the maximum distance from the halo center upto which the remaining
baryons extend. Which of these two ends up being dominant
depends on the details of the actual baryon loss processes.
If one postulates that all halos start 
with at most the 
cosmic baryon fraction ($\sim 0.17$) and that galaxies 
with rising rotation curves are those  that have lost baryons, then 
in these galaxies the smaller sampling of the halo dominates over 
the baryon loss. On the other hand, dwarf spheroidal galaxies which 
have both very large dynamical mass to light ratios (Mateo 1998),
as well as postulated small halo occupation fractions ( Stoehr et al. 2002)
would have to be cases in which the baryon loss dominates over the decrease
in halo occupation. And finally, galaxies with extremely extended gas 
distributions (like NGC~3741) are likely to be those for which there
has been very little net baryon loss, (possibly because the gas has
not been able to collapse to form stars) and the baryons  sample 
a larger fraction of the total halo,

     Which brings us back to the question that why NGC~3741 has been unable to 
convert its baryons into stars. It is often assumed that star formation 
requires a threshold column density.
All the optical emission in NGC~3741 is found to lie within a region with
HI column density $> 1.6\times 10^{21}$ \acc. This makes it very similar to typical
dwarf irregular galaxies, where observationally, one sees that star formation has occurred only
above a threshold column density of $\sim 10^{21}$~\acc (Skillman 1987). While 
this could be a possible explanation as to why the gas in NGC~3741 has not been 
converted to stars, it does not explain why the galaxy has such an unusually 
extended HI disk in the first place. Perhaps galaxies with extended HI disks
are just those which have had a fortuitously long duration disk building history. 
For example the most extended galaxies known  N3714, ESO215-G?009 and 
DDO154 have a tidal index of $\sim$-1.0 (Karachentsev et al. 2004), i.e.
these galaxies are neither so isolated  that the probability of accreting
material is small  nor are they in so dense an environment so as to get perturbed 
(either into a burst of star formation and/or into loosing their gas disks) by 
larger neighbours. A location in a region of modest density enhancement is perhaps 
a necessary (but not sufficient!) condition for a galaxy to quiescently 
and inconspicuously accrete gas into a large disk.

%\begin{acknowledgements}
%We thank the staff of the GMRT that made these observations possible. 
%The GMRT is operated by the National Center for Radio Astrophysics of 
%the Tata Institute of Fundamental Research.
%\end{acknowledgements}

\end{document}